\documentclass[twoside]{article}
\usepackage{fleqn,espcrc2}
\begin{document}


\newcommand{\ttbs}{\char'134}
\newcommand{\AmS}{{\protect\the\textfont2
  A\kern-.1667em\lower.5ex\hbox{M}\kern-.125emS}}

%

\hyphenation{author another created financial paper re-commend-ed Post-Script}


\title{Lepton Number And Neutrino Masses}

\author{Edward Witten\address{School of Natural Sciences,
               Institute for Advanced Study,\\
               Olden Lane, Princeton, NJ 08540, USA}}


\maketitle

\begin{abstract}
I review the arguments that led theoretical physicists in the 1970's to
suspect that lepton number is not an exact symmetry of nature, and
to guess a rough range of neutrino masses.
\vspace{1pc}
\end{abstract}

I should say at the outset that I do not have any real novelty to
offer today.  The study of neutrino masses and neutrino oscillations
is an area where experiment is definitely ahead of theory.  That may
be even more true by the end of this meeting.  All I can do is to
review some ideas about neutrino masses that are not novel and are
undoubtedly familiar to many of you but that remain fresh and
important.
\footnote{This article is based on my lecture at Neutrinos 2000,
Sudbury, Ontario (June 19, 2000).
Because the material is so standard, I have omitted references.
I will just mention one original source where some of the issues
are treated:
S. Weinberg, ``Expectations For Baryon And Lepton Nonconservation,''
First Workshop on Grand Unification (Durham, NH 1980).}

In the 1950's, the discovery of the ``two component'' nature of the
neutrino -- neutrinos  have left-handed spin around their direction
of motion, and antineutrinos have right-handed spin -- was generally
understood to show that the neutrino must be exactly massless.
Indeed, a massive spin one-half particle has two helicity states
(``spin up'' and ``spin down''), while the neutrino only has one.

There is a potential fallacy here, which was understood at an early
stage by a few physicists,\footnote{In the talk after mine, John
Bahcall reminded us of the prescient early work of Bruno Pontecorvo.}
but came to be widely appreciated only in the 1970's in the course of
the development of unified gauge theories of the strong, weak, and
electromagnetic interactions, which are often called GUT's.  In the
course of this work, particle theorists came to widely suspect that
neutrinos have mass and to guess a likely mass range that overlaps
with the range explored in present experiments.

The fallacy is as follows.  To argue, given that the neutrino only has
one helicity state, that it should be exactly massless, one has to
assume that {\it lepton number} is an exactly conserved quantity.  If
lepton number is not conserved, one can treat the left-handed neutrino
and right-handed antineutrino as two different helicity states of one
particle, and combine them to make a massive spin 1/2 particle.  (Such
a particle is called a massive Majorana fermion; this just means that
it is its own antiparticle and in particular carries no conserved
charge.)  If lepton number is an exact symmetry, then the neutrino and
antineutrino, because they have different lepton number, cannot be
combined in this way as two different states of one particle.

In fact, observed particle interactions conserve the baryon number $B$
and the three lepton numbers $L_e$, $L_\mu$, and $L_\tau$.  I will
write $L$ for the total lepton number $L=L_e+L_\mu+L_\tau$.
In the 1970's, theorists generally became convinced that none of these would
be true symmetries, and (hence) that neutrinos would be massive.

I'll try to list, in the rough historical order in which they had
their impact (I depart from the historical presentation at the end),
the principal arguments that suggested that the baryon and lepton symmetries
would be violated.  (Some of the arguments are more compelling for
baryon number.)

(1) All attempts to unify the particles and forces in nature have
required that physicists postulate symmetry-violation, as a result of
putting fermions of different types in the same gauge multiplets.
This was true in the Pati-Salam model, and in the grand unified models
based on groups such as $SU(5)$, $SO(10)$, $E_6$, and it is certainly
true in unified string theories.  To be more exact, the $SU(5)$ model
in its minimal form conserves only one linear combination of the four
symmetries, namely $B-L$, while in $SO(10)$ and $E_6$ (and the string
theories), it is natural to break all of them.  For example, in the
case of the $SU(5)$ model, one combines antiquarks and leptons into a
single gauge multiplet
\begin{equation}
\left(\matrix{\bar q \cr \bar q\cr\bar q\cr\nu\cr e\cr}\right)
\end{equation}
with a somewhat similar structure for the other fermions
As a result, transitions inside the gauge multiplets can violate the
baryon and lepton symmetries.

$B-L$ symmetry is enough to make the neutrinos exactly massless (assuming
that the usual left-handed neutrinos and right-handed antineutrinos are
the only relevant light helicity states).  This is because the neutrino
and anti-neutrino have different values of $B-L$, so the naive argument
for a massless neutrino is valid if $B-L$ is a symmetry.
 Hence in practice the minimal
$SU(5)$ model has massless neutrinos, while most of its more elaborate cousins
(such as the $SO(10)$ and $E_6$ models in which all fermions in a
``generation'' are unified in a single gauge multiplet) have massive
neutrinos.

(2)  At first sight, it might seem that if $B,L_e,L_\mu$, and $L_\tau$
are not fundamental symmetries of nature, there is a mystery to explain:
Why do ordinary particle interactions respect these symmetries?

It turns out that this question has a completely natural answer.
Using the fields of the standard model, it is impossible at the classical
level to violate the baryon and lepton number symmetries by renormalizable
interactions.\footnote{As I explain shortly, $B$ and $L$ violation
in the standard model is induced at an incredibly low level, much too
low to be detectable,   by a quantum anomaly.  The point at the moment
is to explain why ordinary standard model couplings conserve $B$ and $L$.}
Asking why the standard model conserves $B$ and $L$ is thus analogous
to asking why  QED conserves parity even though other forces do not.
The answer seems to be that  QED conserves parity because with only
the  fields of QED (photons and charged leptons), it is impossible to violate
parity by renormalizable interactions.  The same argument explains
conservation of $B,L_e,L_\mu$, and $L_\tau$ by ordinary standard model
processes.

The distinguished role of  renormalizable interactions
comes because they have dimensionless coupling constants
that are pure numbers.  In contrast, unrenormalizable interactions are
expected to be suppressed by a power of $1/M$, where $M$ is a mass scale
at which the standard model breaks down, perhaps a mass $10^{15}$ GeV
suggested by the Georgi-Quinn-Weinberg computation of running of coupling
constants (and hence
characteristic of many grand unified theories), or the Planck mass, roughly
$10^{19}$ GeV, characteristic of gravity.

In the standard model, the lowest dimension operator that violates baryon
number is

\begin{equation}
{1\over M^2}QQQL
\end{equation}
(where $Q$ and $L$ are standard model gauge multiplets containing quark
and lepton fields, respectively) while the lowest dimension operator
that violates lepton number is

\begin{equation}
{1\over M}HHLL,
\end{equation}
where $H$ is a multiplet of Higgs fields.

If we assume that $10^{15}~{\rm GeV}\leq M\leq 10^{19}~{\rm Gev}$,
we get a proton lifetime in the range

\begin{equation}
10^{30}~{\rm years}\leq \tau_{p}\leq 10^{45}~{\rm years}
\end{equation}
and a neutrino mass in the range

\begin{equation}
10^{-5}eV\leq m_\nu \leq 10^{-1}eV.
\end{equation}

For the proton lifetime, there was an attractive model (the minimal
$SU(5)$ model without supersymmetry) that predicted that the proton
lifetime would be quite close to the lower end of the suggested range.  This
prediction has been excluded experimentally, and since then, though I
think most particle theorists still believe that the above estimate is on
the right track,  there has
been no convincing quantitative prediction of $\tau_p$.
For neutrino masses, the considerations have always been qualitative,
and, despite some interesting attempts, there has never been a convincing
quantitative model of the neutrino masses.

What I have explained above is a slightly abstract way to estimate the
neutrino mass.  An elegant mechanism that fits into this general
scheme (and which enables one to see how to make the neutrino mass
bigger or smaller than the above range, if desired) is the ``see-saw
mechanism.''  Here one assumes that a right-handed neutrino $\nu_R$
exists but is very heavy.  In fact, we assume that $\nu_R$ is a
singlet of the standard model gauge group (as the $V-A$ theory of weak
interactions suggests).  One can then have an ordinary Dirac neutrino
mass

\begin{equation}
m\bar\nu_R\nu_L+{c.c.}
\end{equation}
that need not be small.  It originates from a standard model
coupling

\begin{equation}
H\bar \nu_R \left(\matrix{\nu\cr e\cr}\right),
\end{equation}
similar to the couplings that give bare masses to quarks and charged
leptons.  Hence one might expect $m$ to be comparable to the quark
and charged lepton masses.
But $\nu_R$ can also have a very large Majorana mass

\begin{equation}
M\bar\nu_R\bar\nu_R +{c.c.}
\end{equation}
(this violates lepton number conservation, since it couples two
$\bar\nu$'s rather than a $\nu$ and a $\bar \nu$).
When we combine these two terms, we get a $2\times 2$ mass matrix

\begin{equation}
\left(\matrix{ 0 & m \cr m & M\cr}\right)
\end{equation}
for the states $\left(\matrix{\nu_L \cr \bar \nu_R}\right)$.
Supposing that $m<<M$, the result, after ``integrating out'' $\bar\nu_R$, is
a Majorana mass term

\begin{equation}
{m^2\over M}\nu_L\nu_L+{c.c.},
\end{equation}
that is, the ordinary neutrino gets a nonzero $L$-violating mass
$m^2/M$.

If we set $m$ equal to the scale of electroweak symmetry
breaking, and use the above range for $M$,
this leads to the range of neutrino masses suggested above.
Actually, since quark and charged lepton masses vary over many orders
of magnitude (from the electron to the top quark), we  can make the neutrino
mass quite a lot less by making $m$  smaller.  Or we can make the neutrino
masses bigger by making $M$ smaller than the usual GUT scale.  For
example, in conventional GUT theories, this will occur if $\nu_R$ is
naturally massless at tree level.

Whether we take the abstract approach or rely on the see-saw mechanism,
we would expect neutrino oscillations, since it is hard to see why the
neutrino mass matrix would be diagonal in the basis

\begin{equation}
\left(\matrix{\nu_e \cr \nu_\mu \cr\nu_\tau\cr}\right).
\end{equation}

However, large mixing angles seem a bit surprising, at least to me.

(3) So far we have considered arguments that suggest that baryon and
lepton number
violation, and hence neutrino masses, are natural.  Let us now consider
arguments that tend to show that  $B$ and $L$ violation are inevitable.

The first such argument involves electroweak instantons.  't Hooft
showed in 1976 that via instantons of the $SU(2)\times U(1)$ gauge
group, the standard model actually has a mechanism that violates most
of the baryon and lepton number symmetries.  In fact, the only linear
combination of those symmetries that is conserved additively is $B-L$;
in addition, $B$ is conserved mod three (because the standard model
has three generations).  The 't Hooft process will cause the deuteron
to decay to an antiproton plus antileptons at a fantastically slow
rate (the lifetime is far more than $10^{100}$ years).  Despite being
unmeasurably small (and not generating neutrino masses because it
conserves $B-L$), this process shows that, given our present
knowledge, it is all but impossible to interpret $B$ and $L$ as
fundamental symmetries of nature.

Indeed, back in 1976, one could imagine adding additional massive fields to the
standard model in such a way as to cancel the $B$ and $L$
violation by instantons.  With the high precision tests of the standard
model that we now have, this is virtually impossible.

(4) Now I will consider black holes and quantum gravity.  This is the
one point at which I will consider issues that, while certainly not
at all novel, are perhaps not always considered in relation to neutrino
masses.

Classically, a black hole absorbs matter and radiation and does not
emit.  Quantum mechanically, this is impossible as it violates the
hermiticity of the Hamiltonian $H$.  Indeed, if $\langle f|H|i\rangle$
is a nonzero matrix element describing absorption, then its complex
conjugate $\langle i|H|f\rangle$ is a nonzero matrix element
describing emission.  In fact, Hawking showed in 1974 that quantum
black holes do emit approximately thermal radiation,\footnote{The
corrections to the thermal description of the outgoing radiation are,
for a macroscopic black hole, so subtle and small that they have never
been successfully computed.}  at a temperature that is fantastically
small for black holes of astronomical masses.  Such a massive black
hole, placed in empty space, will slowly lose mass to Hawking
radiation.  As it loses mass, it shrinks.  Ultimately, the hole
shrinks down to the Planck scale.  At that point, Hawking's
approximations break down.  By this time, the great bulk of the hole's
mass has been radiated away.  The radiation emitted after Hawking's
approximations break down is presumably highly nonthermal, but can
carry away only a very small fraction of the rest mass of the black
hole, since the bulk of the rest mass has been radiated away during
the period when Hawking's approximations were valid.

Now, let us consider making a black hole from a collapsing star,
or from infalling tables and chairs -- anything with large baryon and lepton
number.  One does not get the baryons back in the outgoing Hawking radiation,
since for the great bulk of the black hole evaporation process, the Hawking
temperature is much too small compared to the proton mass
to enable the emission of any significant
number of baryons or antibaryons.
At the endpoint of the black hole evaporation, we may get back some baryons,
but only a small number since only a small mass remains in the black hole.
So in sum, the formation and evaporation of a black hole does not conserve
baryon number, and hence this cannot be an exactly conserved quantity in
nature.

What about lepton number?  Here we have to be a little more careful,
because it may be that some of the neutrino species are massless, and
if so neutrinos and antineutrinos make up (along with photons,
gravitons, and any other massless particles that may exist in nature)
a large part of the black hole emission.  However, because the black
hole emission is approximately thermal, no significant net lepton
number is emitted during the bulk of the black hole history.  In case
there may exist massless neutrinos in nature, carrying lepton number,
we should worry that perhaps the final pulse of radiation from a black
hole could carry away the original lepton number.  (If so, this
``pulse'' would last a very long time, since Fermi statistics limit
how many neutrinos carrying a bounded energy can be emitted in a given
space in a given time.)  This hypothesis would apparently contradict
subtler aspects of the black hole evaporation story.  Consider a black
hole with a mass $M_S$ comparable to that of the sun.  If the lepton
number of a black hole is a well-defined quantum number, such a black
hole might have a lepton number of $10^{57}$ (if it formed from
collapse of a sun-like star) or any larger number (if the black hole
formed from collapse of a larger mass and then emitted Hawking
radiation to reduce its mass to $M_S$).  So black holes with mass
$M_S$ have infinitely many quantum states, labeled by the lepton
number.  This contradicts the claim that the number of quantum states
of a black hole of given mass is the exponential of the
Bekenstein-Hawking entropy (which is 1/4 of the area of the horizon in
Planck units) and in particular is finite.

I do not claim that arguments such as these are absolutely air-tight,
but they do show that it will be hard to reconcile a claim of absolute
lepton number conservation in nature with what we know about black hole
physics.

In contrast to global symmetries such as baryon and lepton number,
gauge symmetries such as electric charge conservation cause no trouble
for black hole physics.  First of all, electric charge definitely is a
well-defined quantum number of black holes.  One sees it in the
classical (Reissner-Nordstrom) black hole solution; it can be measured
exterior to the black hole horizon, by a flux integral for the
electric field on a large sphere at infinity.  The electric field $E$
of the hole contributes an energy $\int d^3 x\sqrt g\,E^2/8\pi$ to the
black hole mass, so black holes of bounded mass also have bounded
electric charge.  It is {\it global} additive symmetries that cause
difficulties for black hole physics.

Since we have only used general properties of black holes and their
Hawking emission (which is deduced semi-classically and not on the basis
of a detailed microscopic understanding of black holes), the conclusion that
baryon and lepton number are not symmetries of black hole physics
should apply in any consistent theory of quantum gravity.  Hence, we can
test it out in string theory, which is our only real candidate at the moment
for such a theory.  We do not need to assume
that string theory describes nature; the arguments above indicate
that global additive conservation laws such as baryon and lepton number
conservation cannot hold in any consistent quantum gravity theory,
whether it describes nature or not.

We indeed find that in string theory, we never get an additive global
conservation law.  At least in known string vacua, the apparent
additive global symmetries turn out -- sometimes in a tricky fashion
-- to be either gauged or explicitly broken.  For example, one can add
Chan-Paton factors to open strings, apparently introducing a global
symmetry, but upon further examination this turns out to be a gauge
symmetry.  Similarly, in the heterotic string, the global symmetries
on the world sheet turn out to be gauge symmetries (not global
symmetries) in space time.  As an example of the opposite kind, in
compactification of the Type II or heterotic string theories from ten
to four dimensions on a compact manifold $K$, one gets approximately
massless ``axions'' associated with the second Betti number $b_2(K)$.
The shift symmetries of these axions appear to be continuous global
symmetries (albeit spontaneously broken), but on closer examination,
one finds that these symmetries are explicitly violated by world-sheet
instantons.

In summary, these arguments seem to show that $B$ and $L$ cannot
be conserved quantities in any theory that includes quantum gravity.
This is a strong hint of neutrino masses, but there is a loophole.  Although
$L$, for example, cannot be an exactly conserved quantity, it might be
conserved mod 7, or mod $n$ for any integer $n$.  For $n>2$, this would
suffice to prevent neutrino masses.  If    one wishes in string theory
to make a model in which neutrino masses are exactly zero, one would
arrange to have a discrete ${\bf Z}_n$ symmetry producing such a mod $n$
conservation law.  Discrete symmetries that prevent neutrino masses
could well prevent quark or charged lepton masses, so it actually would
take some care to arrange a model in which the neutrinos are exactly
massless and the other fermions have mass.

(5) Finally, the last hint that $B$ and $L$ are not exact symmetries
in nature is simply that the real universe has a small but nonzero $B$
and $L$ density relative to the entropy density. (In the case of $L$,
we do not know this for sure, since the observed net density of
electrons might conceivably be canceled by not yet observed relic
antineutrinos.)  If $B$ and $L$ are not exact symmetries, then making
use of the $C$ and $CP$ violation in particle interactions, and the
$CPT $ violation in cosmology that comes from the fact that the
universe is expanding, it is possible to spontaneously generate
nonzero $B$ and $L$ densities.  There is not a convincing quantitative
theory, but the observed $B$ and $L$ densities (about $10^{-9}$ of the
entropy density) are in a natural range.  If $B$ and $L$ are exactly
conserved quantities, they must be postulated as initial conditions in
the Big Bang; this is certainly less attractive.

I hope that today  I have at least managed, albeit without saying anything
that is really new, to convey a sense of the exciting and wide-ranging
theoretical ideas relevant to the neutrino mass question.  Hopefully,
experimentalists are beginning to come back with answers.  I am sure we
will hear more in the next few days.

\section*{Acknowledgments}
This work was supported in part by NSF Grant PHY-9513835 and the
Caltech Discovery Fund. On leave at Department of Physics, Caltech,
Pasadena CA 91125.



\end{document}